\documentclass[prl,superscriptaddress,twocolumn]{revtex4}

\usepackage{amsfonts,amssymb,amsmath}
\usepackage[]{graphics,graphicx,epsfig}
\usepackage{amsthm}

\def\identity{\leavevmode\hbox{\small1\kern-3.8pt\normalsize1}}

\renewcommand{\epsilon}{\varepsilon}

\begin{document}
\title{Entropic test of quantum contextuality}

\date{\today}
\author{P. \surname{Kurzy\'nski}}
\affiliation{Centre for Quantum Technologies, National University of Singapore, 3 Science Drive 2, 117543 Singapore, Singapore}
\affiliation{Faculty of Physics, Adam Mickiewicz University, Umultowska 85, 61-614 Pozna\'{n}, Poland} 
\author{R. \surname{Ramanathan}}
\affiliation{Centre for Quantum Technologies, National University of Singapore, 3 Science Drive 2, 117543 Singapore, Singapore}
\author{D. \surname{Kaszlikowski}}
\email{phykd@nus.edu.sg}
\affiliation{Centre for Quantum Technologies, National University of Singapore, 3 Science Drive 2, 117543 Singapore, Singapore}
\affiliation{Department of Physics, National University of Singapore, 2 Science Drive 3, 117542 Singapore, Singapore}

\begin{abstract}
We study the contextuality of a three-level quantum system using classical conditional entropy of measurement outcomes.  First, we analytically construct the minimal configuration of measurements required to reveal contextuality. Next, an entropic contextual inequality is formulated, analogous to the entropic Bell inequalities derived by Braunstein and Caves in [Phys. Rev. Lett. {\bf 61}, 662 (1988)], that must be satisfied by all non-contextual theories. We find optimal measurements for violation of this inequality. The approach is easily extendable to higher dimensional quantum systems and more measurements. Our theoretical findings can be verified in the laboratory with current technology.   
\end{abstract}





\maketitle

{\it Introduction} In information theory the relation between two events can be quantified using the notion of conditional entropy. This notion was successfully applied to study the relation between space-like separated measurements on different parts of an entangled quantum system and it was shown that conditional entropies for measurements on such systems do not obey classical properties of entropy \cite{BC}, which is yet another manifestation of quantum nonlocality \cite{EPR}. Since quantum nonlocality is only a special case of quantum contextuality \cite{KS} it is natural to ask whether classical properties of entropy also fail do describe measurements in contextual scenarios.  

The notion of contextuality, as introduced by Kochen and Specker (KS) \cite{KS}, can be explained as follows. Suppose that a measurement $A$ can be jointly performed with either $B$ or $C$. Measurements $B$ and $C$ are said to provide a context for the measurement $A$. The measurement $A$ is contextual if its outcome does depend on whether it was performed together with $B$ or with $C$. Therefore, the essence of contextuality is the lack of possibility to assign an outcome to $A$ prior to its measurement and independently of the context in which it was performed. The crucial observation by KS \cite{KS} was that quantum theory is contextual for any system whose dimension is greater than two. The seemingly different Bell theorem \cite{Bell} is in fact a special instance of the KS theorem where contexts naturally arise from the spatial separation of measurements. 

In mathematical terms quantum nonlocality and contextuality can be formulated in terms of probability theory. Specifically, the reason behind both Bell and KS theorems is the lack of joint probability distribution for all measured observables \cite{Fine,Abramsky,Liang}. For example, imagine some physical system on which one can perform various measurements denoted 
by $A_1, A_2,\dots, A_N$. Each measurement $A_i$ yields an outcome $a_{i, j_i}$ (where $j_i$ enumerates outcomes) with probability $p(A_i = a_{i, j_i})$. The non-contextuality hypothesis is true if and only if there exists a joint probability distribution for the outcomes of all observations, i.e., $p(A_1 = a_{1, j_1}, A_2 = a_{2, j_2},\dots, A_N = a_{N, j_N})$, such that one can recover all the measurable probabilities as its marginals. For instance, consider that the subset of measurements $\{A_{k_1} , \dots, A_{k_l}\}$ with $\{k_1, \dots, k_l\} \subseteq \{1, \dots, N\}$  can be jointly performed, in other words, the probability distribution $p(A_{k_1} = a_{k_1, j_{k_1}}, \dots, A_{k_l} = a_{k_l, j_{k_l}})$ can be experimentally determined. The non-contextuality hypothesis then requires that the joint probability distribution for all measurements,  $p(A_1 = a_{1, j_1}, A_2 = a_{2, j_2},\dots, A_N = a_{N, j_N})$ recovers any such $p(A_{k_1} = a_{k_1, j_{k_1}}, \dots, A_{k_l} = a_{k_l, j_{k_l}})$ as its marginal, i.e., 
\begin{eqnarray}
&&p(A_{k_1} = a_{k_1, j_{k_1}}, \dots, A_{k_l} = a_{k_l, j_{k_l}}) = \nonumber \\ 
&&\sum_{\bar{a}_{k, j_k}}p(A_1 = a_{1,j_1},\dots, A_N = a_{N,j_N}). \nonumber
\end{eqnarray}
Here the summation is over the outcomes of all the measurements $A_j$ that are not in the jointly measurable subset.
From hereon, we denote probabilities by $p(A_i)$ instead of $p(A_i = a_i)$ for notational convenience, wherever there is no possibility of confusion.

An important question is the minimal number of measurements on some quantum system that one has to perform in order to observe contextuality and prevent the existence of their joint probability distribution. The currently known most economic proofs for a three-dimensional system (qutrit) consist of 5 measurements in case of a state-dependent test \cite{Kl} and of 13 measurements in case of a state-independent test \cite{Yu}. Qutrits are of special interest since they are not only the smallest contextual systems, but they also physically correspond to a single system to which the concept of nonlocality and entanglement cannot be unambiguously applied. Therefore, a single qutrit together with the most economic set of contextual measurements can be considered as a primitive of quantum contextuality in a similar sense as an entangled pair of qubits together with the CHSH (Clauser-Horn-Shimony-Holt) inequality \cite{CHSH} is considered as a primitive of quantum nonlocality.

In this Letter we analytically find the minimal number of contextual measurements for a single qutrit. A new graph theoretical method to construct joint probability distributions is presented and it is proven that for sets of measurements whose corresponding graphs do not contain cycles (i.e., there are no subsets of measurements that cyclic-commute) there always exists a joint probability distribution reproducing quantum marginals. Next, using the minimal set of measurements we formulate an entropic contextual inequality in the spirit of \cite{BC}. Finally, we find measurement settings for an optimal violation of this inequality for a single qutrit. 

{\it Minimal number of contextual measurements} In order to find the minimal number of contextual measurements one is tempted to start with two measurements $A$ and $B$. However, this does not work because (i) either $A$ and $B$ commute in which case quantum mechanics itself provides a joint probability distribution, or (ii) $A$ and $B$ do not commute and one can simply write $p(A,B) = p(A)p(B)$, a joint probability that reproduces the marginal probabilities $p(A)$ and $p(B)$. Observe that a single particle in one dimension for which measurements of position and momentum do not commute is not contextual because of (ii). Next, consider three measurements: $A$, $B$ and $C$. The various scenarios are as follows: (i) All three measurements mutually commute, which is equivalent to (i) for two measurements (ii) All of them do not commute, which allows us to define $p(A, B, C) = p(A)p(B)p(C)$ (iii) Only one pair commutes ($A$ and $B$) in which case the joint probability distribution is $p(A, B, C) = p(A, B)p(C)$, where $p(A, B)$ is provided by quantum mechanics. (iv) One pair of them ($B$ and $C$) does not commute in which case one may construct $p(A, B, C) = p(A, B)p(A, C)/p(A).$ This joint probability distribution reproduces all measurable marginals, therefore the system that has only two contexts is not sufficient to refute non-contextuality. The next case of four measurements was shown to be sufficient to prove this discrepancy for a system of dimension at least four, and is known as the CHSH inequality \cite{CHSH}.

\begin{figure}
\scalebox{0.5}
{\includegraphics{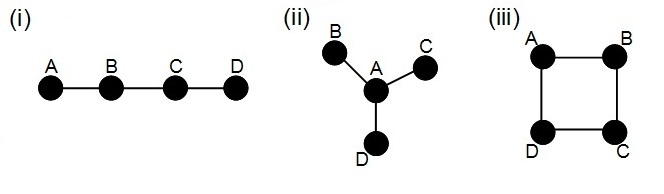}}
\vspace{-0.5 cm}
\caption{\label{f1} Graphical notation of commutation relations. Vertices of the graph correspond to observables and edges represent commutativity.}
\end{figure}

Can we show the discrepancy for a three-level system and only four measurements?
To show that the answer is no, it is convenient to introduce graphic notation as in Fig. \ref{f1}.
Each vertex represents a measurement and edges represent commutation between the connected measurements. The only significant scenarios that do not reduce to previous considerations are represented by the chain graph (in Fig. \ref{f1} (i)), the star graph (ii) and the cycle (iii). For (i) we construct $p(A,B,C,D)=p(A,B)p(B,C)p(C,D)/(p(B)p(C))$, for (ii) $p(A,B,C,D)=p(A,B)p(A,C)p(A,D)/(p(A)p(A)).$ Note that the probabilities on the right-hand side of these equations exist due to the assumption of joint measurability. Measurements corresponding to the graph in Fig. \ref{f1} (iii) do not exist for a three-level system. This is because in order to have $[A,B]=0$ and $[A,D]=0$, but $[B,D]\neq 0$ one requires $A$ to be a degenerate operator. In the case of a three-level system, this means that two eigenvalues of $A$ are the same and therefore, without loss of generality $A$ can be set to be a projector of rank one. Therefore all four measurements $A$, $B$, $C$ and $D$ are rank one projectors. The cycle graph (iii) implies that both $A$ and $C$ are orthogonal to $B$ and $D$. Since we require that $B \neq D$, these two projectors span a plane orthogonal to both $A$ and $C$ which in three-dimensional Hilbert space implies $A = C$. The problem then reduces to case (iii) for three measurements.

\begin{figure}
\scalebox{0.5}
{\includegraphics{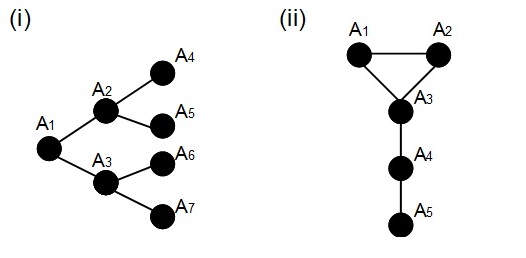}}
\vspace{-0.5 cm}
\caption{\label{f2} Two examples of graphs of observables admitting joint probability distributions.}
\end{figure}

For three-level systems one requires at least five projective measurements to show the lack of joint probability distribution. Before we proceed, let us prove one property of the construction used above, namely that for any commutation graph which does not contain cycles (tree graph) there always exits a joint probability distribution consistent with quantum theory. This construction is given by the product of probability distributions corresponding to the edges of the graph (denoted by the set $E(G)$) divided by the product of probabilities of common vertices, where a vertex $i \in V(G)$ (the set of vertices of the graph) of degree $d(i)$ (the number of nearest neighbors) appears $d(i)-1$ times in the product, i.e., $$p(A_1,\dots,A_N)=\frac{\prod_{(i,j)\in E(G)}p(A_i, A_j)}{\prod_{i\in V(G)}p(A_i)^{d(i)-1}}.$$ Since quantum theory provides joint probability distributions for any two commuting observables, this construction recovers any measurable marginal as can be seen by summing over all other observables, starting the summation from the leaves (free ends of the tree). For example, for the instance presented in Fig. \ref{f2} (i) the joint probability distribution is 
\begin{eqnarray}
p(A_1,\dots,A_7)&=&\frac{p(A_1, A_2)p(A_1, A_3)p(A_2, A_4)p(A_2, A_5)}{p(A_1)p(A_2)^2}  \nonumber \\  &\times & \frac{p(A_3, A_6)p(A_3, A_7)}{p(A_3)^2} \nonumber
\end{eqnarray}
and for instance to recover $p(A_2, A_5)$ the summation order is $A_7,A_6,A_4,A_3,A_1$. In fact, the construction can also be applied to the scenario when the commutation graph apart from the tree structure also contains cliques (a clique is a fully conntected subgraph which by definition is a jointly measurable subset).

Since all open graphs are trees for which joint probability distribution exists and for a three-level system one requires at least five projective measurements, the minimal graph for which one can show the discrepancy is a pentagon (5-cycle). For other graphs with cycles smaller than five, such as the example in Fig. 2 (ii), one can always find joint probability distributions, for example $$p(A_1,\dots,A_5)=\frac{p(A_1,A_2,A_3)p(A_3,A_4)p(A_4,A_5)}{p(A_3)p(A_4)}.$$ The similar case with a square (4-cycle) does not work due to reasons already discussed. This analytic result confirms observation in \cite{Kl} that projectors corresponding to the 5-cycle are necessary and sufficient to reveal the contextuality of a single three-level system. 

{\it Entropic contextual inequality} We now focus on a three-level system which can physically correspond to spin 1. Spin states are represented by rank 1 projectors $A_{n}^s$, where $s=0,\pm 1$ and $n$ denotes the direction of spin projection. We say that $A_{n}^s$ is measured if it is an eigenprojector of the measured observable. It is well known that the squares of spin 1 operators $S_n^2$ and $S_m^2$ for two orthogonal directions $n$ and $m$ can be jointly measured. An example of the observable corresponding to this measurement is $S_n^2 - S_m^2$. We say that this observable provides the context $\{A_{n}^0,A_{m}^0\}$. However, the same operator $S_n^2$ can be jointly measured with any other operator $S_{m'}^2$, where $m'$ is confined to the plane orthogonal to $n$, hence there are many contexts for the measurement of $A_n^0$.

Let us derive an entropic contextual inequality analogous to the entropic Bell inequality in \cite{BC}. It involves five projectors $\{A_1,A_2,\dots,A_5\}$ ($A_i=|A_i \rangle\langle A_i|$) on which we impose cyclic orthogonality restrictions, i.e. $A_i A_{i+1}=0$, where the subscript is modulo five. Neighboring projectors are jointly measurable since they are orthogonal. As a result, for every projector $A_i$ there exist two contexts $\{A_i,A_{i\pm1}\}$. In case of spin 1 these projectors may correspond to $A_i^0$, where $i$ and $i\pm1$ denote orthogonal directions in real space. Let us start with the assumption that despite the fact that not all projectors are jointly measurable, there exists a joint probability distribution for all five projectors $p(A_1,\dots,A_5)$. This joint probability distribution is a non-contextual description of the measurements $\{A_i\}$. It is then possible to define the joint entropy $H(A_1,A_2,A_3,A_4,A_5)$, where $H(A) = -\sum_a p(A=a)\log p(A=a)$ denotes the Shannon entropy. Two classical properties of the Shannon entropy are used in the derivation of the entropic contextual inequality as in \cite{BC}. The first is the chain rule  $H(A,B)=H(A|B)+H(B)$ and the second is $H(A|B) \leq H(A) \leq H(A,B)$. The latter inequality has the intuitive interpretation that two random variables cannot contain less information than one of them and that conditioning cannot increase information content of $A$. The conditional entropy above is defined as 
\begin{equation}
H(A|B)=\sum_{b=0,1}p(B=b) H(A|B=b), \nonumber
\end{equation} 
where the entropy $H(A|B=b)$ is defined like standard entropy, but using conditional probabilities $p(A=a|B=b)$. The conditional entropy $H(A|B)$ describes the information content of $A$ given the value of $B$. 

Repeated application of the chain rule yields
\begin{eqnarray}
& &H(A_1,A_2,A_3,A_4,A_5) = H(A_1|A_2,A_3,A_4,A_5)+ \nonumber \\ 
& &H(A_2|A_3,A_4,A_5)+H(A_3|A_4,A_5) + H(A_4|A_5)+ H(A_5). \nonumber 
\end{eqnarray}
Using the inequality $H(A|B) \leq H(A) \leq H(A,B)$, one then obtains the entropic contextual inequality
\begin{equation}\label{e3}
H(A_1|A_5)\leq H(A_1|A_2)+H(A_2|A_3)+H(A_3|A_4) + H(A_4|A_5).
\end{equation}
\begin{figure}
\scalebox{0.5}
{\includegraphics{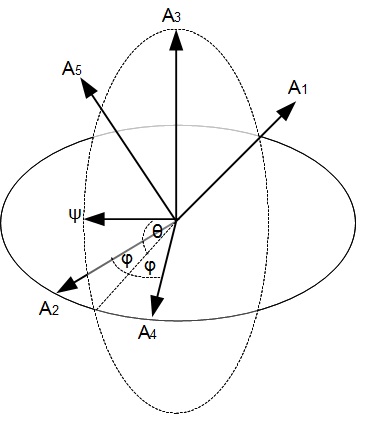}}
\caption{\label{f3} Configuration of projectors leading to maximal violation of the entropic contextual inequality for three-level systems.}
\end{figure}
We define the quantity 
\begin{equation}
{\cal{C}}=H(A_1|A_5) - H(A_1|A_2) - H(A_2|A_3)-H(A_3|A_4) - H(A_4|A_5), \nonumber
\end{equation}
therefore the inequality (\ref{e3}) can be rewritten as ${\cal{C}}\leq 0$.
For the three-level system the maximal violation of this inequality can be shown to be of magnitude $0.091$ bits. The optimal solution can be written as follows with parameters $\theta=0.2366$ and $\varphi=0.1698$ 
\begin{figure}
\scalebox{0.7}
{\includegraphics{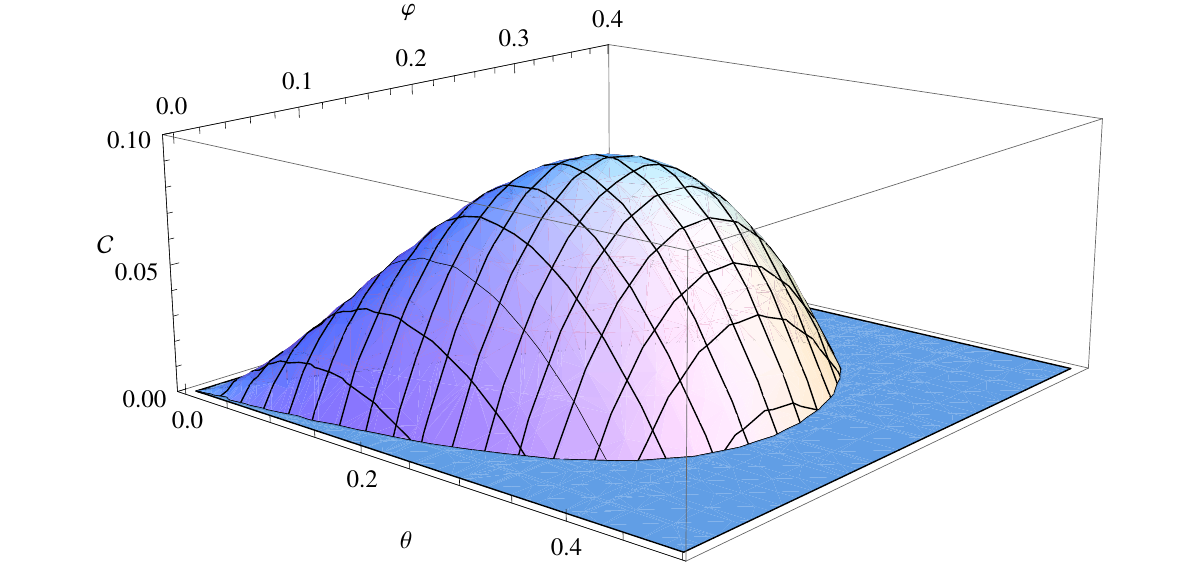}}
\caption{\label{f4}  The plot of positive part of ${\cal{C}}$ as a function of $\theta$ and $\varphi$. Maximal violation of the entropic contextual inequality is seen to be $0.091$ bits.}
\end{figure}
(see Fig. \ref{f4} where we plot ${\cal{C}}$ as a function of $\theta$ and $\varphi$)
\begin{eqnarray}
& &|\psi\rangle  =  (\sin\theta,  \cos\theta, 0)^T,~|A_1\rangle = \left(\frac{\sqrt{\cos 2\varphi}}{\sqrt{2}\cos\varphi}, \frac{\tan\varphi}{\sqrt{2}}, \frac{1}{\sqrt{2}}\right)^T, \nonumber \\
& &|A_2\rangle = (0, \cos\varphi, -\sin\varphi)^T,~|A_3\rangle = (1, 0, 0)^T, \nonumber \\
& &|A_4\rangle = (0,  \cos\varphi, \sin\varphi)^T,~
|A_5\rangle = \frac{|A_1\rangle \times |A_4\rangle}{||~|A_1\rangle \times |A_4\rangle ~||}, \nonumber 
\end{eqnarray}
where $\times$ denotes the three-dimensional cross product. The projectors corresponding to $|A_1\rangle$ and $|A_5\rangle$, in addition to orthogonality obey the symmetries (i) $\langle A_5|\psi\rangle = \langle A_1|\psi\rangle$, (ii) $\langle A_5|A_2\rangle = \langle A_1|A_4\rangle$ and (iii) $\langle A_5|A_3\rangle = \langle A_1|A_3\rangle$. These symmetries uniquely define $|A_1\rangle$ and $|A_5\rangle$.

The intuitive reason for the appearance of these symmetries in the optimal solution is the following. Maximal violation of the entropic contextual inequality requires maximizing $H(A_1|A_5)$ while simultaneously minimizing the right-hand side of the inequality. For orthogonal projectors $A$ and $B$ one has $H(A|B) = p(B=0) H(A|B=0)$. If $A$, $B$ and the state $|\psi\rangle$ are coplanar, $H(A|B) = 0$ because then $p(A=0) = p(B=1)$ and $p(A=1) = p(B=0)$. Therefore, we need to set all pairs of projectors corresponding to entropies on the right-hand side of the inequality as coplanar with $|\psi\rangle$ as possible, whilst maximizing $H(A_1|A_5)$. The symmetries listed above arise as a consequence of these considerations. Furthermore, numerical optimization over the five projectors and the state also reveals these symmetries for the solution. Note that any pure state of a three-level system violates inequality (\ref{e3}), the optimal projectors being obtained from the above solution by Euler rotations.

{\it Discussion} It is important to notice that all entropies in the inequality (\ref{e3}) can be evaluated within quantum theory since they refer to jointly measurable quantities. Although the entropic inequality constructed here involves five projectors as in \cite{Kl}, it is not equivalent to the pentagram inequality constructed there. For the pentagram inequality, violation is obtained if and only if the joint probability distribution does not exist. Violation of the entropic contextual inequality (\ref{e3}) therefore implies violation of the pentagram inequality but the converse is not true. The optimal projectors for violation of (\ref{e3}) do not possess the symmetry of the projectors in the pentagram. For the optimal projectors and state given above, the violation of the pentagram inequality is $0.049$ which is less than the maximal value of $\sqrt{5}-2$. The reason for the asymmetry of optimal projectors in the inequality (\ref{e3}) and in the Fig. \ref{f3} is that the projectors $A_1$ and $A_5$ are special in the sense that $H(A_1|A_5)$ has to be maximized, whereas $H(A_i|A_{i+1})$ for $i=1,\dots,4$ has to be minimized. On the other hand, the pentagram inequality projectors are treated on equal footing. 

The pentagram inequality has been recently tested in the laboratory in \cite{Zeilinger}, a similar setup can be used to test the entropic inequality as well. Entropic contextual inequalities can be easily constructed for more projectors than five and applied to higher dimensional quantum systems following the construction above. 
Since these inequalities are not equivalent to those following from the approach in \cite{Kl}, an interesting problem is to investigate the set of quantum states that violates entropic inequalities as opposed to the set that violates the inequalities in \cite{Kl}. The earlier approach is based on studying the extremal edges of a polyhedral cone, which leads to a finite set of inequalities that are hard to construct and interpret. Entropic contextual inequalities are simpler to construct and carry a clear information-theoretic interpretation. The violation of the entropic contextual inequality indicates that the joint probability distribution does not exist. Insistence on a joint probability distribution would result in negative information whose deficit is measured by the violation of the inequality. Moreover, for a single three-level system no entanglement exists and therefore violation of the entropic inequality is solely due to contextuality, unlike the entropic Bell inequality in \cite{BC} where entanglement was necessary. It is interesting how these inequalities extend to macroscopic systems where entropies arise naturally in the context of thermodynamics. 

{\it Conclusions} In this paper we have constructed an entropic contextual inequality that can be applied to the simplest indivisible quantum system, namely a single three-level system. After analytically showing that the minimal commutation graph for this system is the 5-cycle confirming earlier observations, we constructed the optimal set of projectors that maximally violate this inequality. Note that a different information-theoretic approach to contextuality has been considered in \cite{Stephanie} using the notion of min-entropy. Also, the constraints on the Shannon entropies of marginal probabilities from the existence of a joint probability distribution for graphs that are n-cycles has been considered in \cite{Fritz,FC}. The construction of the contextual inequalities considered in this paper can be easily extended to other measures of disorder that obey the intuitive classical properties used in the proof. The entropic Bell inequality derived in \cite{BC} was generalized in \cite{CA} using the notion of mutual information, therefore an open question is to investigate which information-theoretic quantities optimally reveal the lack of objective realism in quantum systems. A further question is to fully analyze the relation between the entropic contextual inequality derived here and the pentagram inequality. It would also be interesting to find optimal commutation graphs that reveal contextuality for given system dimensions \cite{Budroni}. This work is supported by the National Research Foundation and Ministry of Education in Singapore.

\end{document}